\documentclass[aps,prd,reprint,showpacs,preprintnumbers,nofootinbib]{revtex4-1}

\usepackage{graphicx,amsmath,amssymb}

\newcommand{\cl}{\textsc{l}}
\newcommand{\ccr}{\textsc{r}}
\newcommand{\cd}{\textsc{d}}
\newcommand{\lp}{\lambda_+}
\newcommand{\lnn}{\lambda_-}
\newcommand{\bea}{\begin{eqnarray}}
\newcommand{\eea}{\end{eqnarray}}
\newcommand{\nn}{\nonumber}
\newcommand{\ov}{\overline}
\newcommand{\ang}[1]{\langle #1 \rangle}

\mathchardef\mhyphen="2D

\def\a{\alpha}

\def\g{\gamma}

\def\e{\epsilon}

\def\th{\theta}

\def\l{\lambda}
\def\m{\mu}
\def\n{\nu}

\def\p{\pi}

\def\s{\sigma}
\def\t{\tau}

\def\f{\phi}
\def\x{\chi}

\def\G{\Gamma}
\def\D{\Delta}

\def\W{\Omega}

\begin{document}

\title{A Model For Late Dark Matter Decay}

\author{Nicole F. Bell}
\affiliation{School of Physics, The University of Melbourne, Victoria 3010, Australia}

\author{Ahmad J. Galea}
\affiliation{School of Physics, The University of Melbourne, Victoria 3010, Australia}

\author{Raymond R. Volkas}
\affiliation{School of Physics, The University of Melbourne, Victoria 3010, Australia}

\date{\today}

\begin{abstract}

The standard cold dark matter cosmological model, while successful in
explaining the observed large scale structure of the universe, tends
to overpredict structure on small scales.  It has been proposed this
problem may be alleviated in a class of late-decaying dark matter
models, in which the parent dark matter particle decays to an almost
degenerate daughter, plus a relativistic final state.  We construct explicit particle physics models that realize this goal while
obeying observational constraints.  To achieve this, we introduce a
pair of fermionic dark matter candidates and a new scalar field, which
obey either a $\mathbb{Z}_4$ or a $U(1)$ symmetry.  Through the
spontaneous breaking of these symmetries, and coupling of the new
fields to standard model particles, we demonstrate that the desired
decay process may be obtained.  We also discuss the dark matter
production processes in these models.
\end{abstract}


\maketitle

\section{Introduction}

There is an abundance of evidence to indicate the existence of dark
matter (DM), including its necessary contribution to both galactic
stability and structure formation in the early
Universe~\cite{Bertone:2004pz,Jungman:1995df,Liddle:1993fq}.  The
standard $\Lambda$CDM cosmological model, in which cold dark matter
(CDM) makes up 22\% of the universal energy budget, provides an
excellent description of our Universe.  However, little is known about
the particle properties of dark matter.  In addition, some problems
with CDM are encountered at small scales.

A popular class of CDM candidates is weakly interacting massive
particles (WIMPs). In WIMP models the DM couples weakly to standard
model (SM) particles, which allows for scattering/annihilation
processes.  These serve to keep the dark sector in thermal equilibrium
with the visible sector in the early Universe and can, with an
appropriate choice of coupling, cause the DM to freeze out with the
correct relic density.  Interaction with the SM is similarly appealing
from a detection standpoint, potentially providing both
direct~\cite{Ahmed:2008eu,Angle:2007uj,Bernabei:2008yi} and
indirect~\cite{Adriani:2008zr,Ackermann:2010rg,Knodlseder:2005yq}
signatures of a given model. Nonobservation of these signatures
allows for constraints to be placed on parameters such as particle
masses or coupling constants.

Though the DM mass is unknown, some information can be inferred from
observations of large scale structure.  For cold dark matter,
structure forms hierarchically, with the earliest structures formed on
short length scales, which can then merge to form larger structures.
This is to be contrasted with hot dark matter in which the largest
superclusters form first.  Numerical simulation has shown the CDM
scenario to fit observations well~\cite{Diemand:2006ik}, while hot
dark matter is strongly disfavored.

The CDM model is not-problem free, however, as it tends to overproduce
small scale power
\cite{Diemand:2006ik,Moore:1999gc,Navarro:2003ew,Gentile:2004tb,Salucci:2007tm,Diemand:2006ik,Gilmore:2006iy,Gilmore:2007fy,Gilmore:2008yp,Wyse:2007zw,Governato:2002cv,Klypin:1999uc,Metcalfe:2000nv,Peebles:2001nv,SommerLarsen:1999jx,Tikhonov:2009jq}. Simulations
predict cusps in the DM density at the centers of galactic halos in
conflict with observation.  CDM also over-predicts the number of dwarf
galaxies orbiting a Milky Way-sized galaxy by about a factor of
$10$. Although simulations do not include visible matter, the
gravitational potential wells they predict would promote a level of
star formation not observed.  Though these issues may be partially
alleviated by tidal disruption and other effects, the small scale
power problems of $\Lambda$CDM are still poorly understood (see
e.g.~\cite{D'Onghia:2009pz,Schneider:2010jr} for recent work). Such
issues have led many to consider a ``warm'' DM candidate, with a mass of
keV scale, intermediate between hot and cold dark matter.  In this
work, as in
\cite{Finkbeiner:2007kk,ArkaniHamed:2008qn,TuckerSmith:2001hy,Chang:2008gd,Zurek:2008qg,Feldman:2010wy,Profumo:2009tb,Cline:2010kv},
we will consider an alternative hypothesis in which the usual
assumption of a single DM candidate is challenged.

We consider a scenario with two WIMP candidates, in which one species
is unstable to decay into the other.  If the mass splitting between
the two WIMPS is sufficiently small, the decay process will leave the
overall halo mass unaffected, while giving its constituent DM
particles a small velocity kick.  Such velocity kicks heat the dark
matter halos and cause them to expand, softening the central cusps and
disrupting small
halos~\cite{Melia,Peter:2010jy,Cembranos:2005us,Kaplinghat:2005sy,SanchezSalcedo:2003pb}.
Such models are appealing, as they can alleviate the small scale
structure problems, while retaining the attractive features of cold
dark matter.

We shall assume the DM decays predominantly
via the channel
\bea
\x^*\rightarrow\x+l\,,\label{dec}
\eea
where $\x^*$ and $\x$ denote the heavier and lighter candidate, respectively, and $l$ is some relativistic final state. The mass splitting between $\x^*$ and $\x$ is given by 
\bea
\D m=m_{\x^*}\e\,,\label{delm}
\eea
where $\e\ll1$. 
Abdelqader and Melia~\cite{Melia} have shown the dwarf halo problem can be solved 
for $\e\simeq(5-7)\times10^{-5}$ and a decay lifetime of $(1-30)$ Gyr.
The work of Peter, Moody, and Kamionkowski~\cite{Peter:2010jy} has
demonstrated that galaxy cusps can be alleviated for a wider range
of $\e$ and $\t_{\x^*}$, with the most favored lifetimes in the range
$(0.1-100)$ Gyr. Subsequent work by Peter and
Benson~\cite{Peter:2010sz} has used properties of galactic subhalos
to further constrain the allowed values of $\e$, preferring lower
values to those favored in \cite{Melia}. Dark matter decays may be further constrained from analysis of their effect on weak lensing of distant galaxies as in~\cite{Wang:2010ma}. However, at present such analyses have only placed limits on models with much larger values of $\e$ than those considered in this work.

An interesting possibility, from an observational standpoint, is a
decay mode in which the relativistic final state, $l$, consists of SM
particles.  This allows the possibility of verifying the model, via the
detection of particles produced by decay in our own Galaxy, or of a
diffuse flux from decays in halos throughout the Universe.  Current
astrophysical observations place constraints on the allowed
parameters, via comparison of the decay fluxes with relevant
astrophysical backgrounds.  Reference~\cite{Yuksel:2007dr} placed stringent
constraints on the decay parameters for the case in which $l$ is a
photon, while Ref.~\cite{Bell:2010fk} derived somewhat weaker
constraints for the cases in which $l\,=\overline{\nu}\nu$ or $e^\pm$.
For $l\,=\gamma$ or $e^\pm$ the lifetime is restricted to be below about
1 Gyr, while a much larger range of lifetimes is permitted for
$l\,=\overline{\nu}\nu$ .

The aim of this work is to construct a particle physics model which
can realize the decaying dark matter scenario. We shall use the
criterion specified by Abdelqader and Melia \cite{Melia} [namely,
$\e\simeq(5-7)\times10^{-5}$ and $\t\sim(1-30)$ Gyr] as a reference
point for these models, but given the constraints of
Ref.~\cite{Peter:2010sz}, we will choose the more restrictive value
of $\e\sim10^{-5}$ and $\t\simeq(1$-$10$) Gyr. In
Sec.~\ref{sec:models} we introduce and discuss two possible models
for decaying dark matter, and outline the DM production mechanism.
Section~\ref{sec:constraints} focuses on constraints on the models and
the available regions of parameter space. We conclude in
Sec.~\ref{sec:conclude}.

\section{A Dark Matter Decay Model}
\label{sec:models}

In order to construct a model which can achieve this decay scenario there are certain criteria that need to be satisfied. The first and  most important of these is the need for two candidates with nearly degenerate masses. Second, we need either decay of the parent DM particle to light SM final states, or to some new light degree of freedom. Third, the process needs to occur on times scales relevant for the disruption of structure formation, and last, we need some viable DM production mechanism. WIMP-like scenarios are particularly interesting on this front, as WIMPs are populated as thermal relics and naturally freeze out in the early Universe with the correct relic density.

Two scenarios will be considered. In the first we implement a variation of the ``exciting dark matter'' model conceived by Finkbeiner and Weiner \cite{Finkbeiner:2007kk}, which involves the addition of a dark sector containing a Dirac fermion and a real scalar field to the standard model. The introduced fields obey a discrete $\mathbb{Z}_4$ symmetry, the breaking of which leads to a nondegeneracy of the masses of the fermion's two Weyl components, and an instability of the heavier to decay into the lighter. The scenario considered in this work differs from \cite{Finkbeiner:2007kk} in the values of the model parameters chosen; in short, we consider longer decay lifetimes. 

As a second example, we consider a generalization of the scenario in \cite{Finkbeiner:2007kk}, in which we replace the $\mathbb{Z}_4$ symmetry with a global $U(1)$ symmetry, requiring the introduced scalar field to be complex. The breaking of this $U(1)$ will produce a pseudo-Nambu-Goldstone boson, which will serve as our light final state for the dominant decay channel. We show that production through interaction with the SM is impossible in the $U(1)$ model. The second scenario is one example among any number of generalizations and extensions to the simple $\mathbb{Z}_4$ model; it is simply an illustration that decaying DM can be realized in a particle model irrespective of the strength of coupling to the SM.

In Sec.~\ref{Z4} we shall explore a model in which SM final states are produced in the dominant decay channel. In \ref{U1} we consider the possibility of completely non-SM final states. In \ref{production} we discuss production, and in \ref{depopulation} consider the possibility of $\x^*$ depopulation.

\subsection{SM Final States $(\mathbb{Z}_4)$}
\label{Z4}

When searching for a light final state for the process in Eq.(\ref{dec}) the obvious place to look is the SM, as the existence of particles with masses substantially below the CDM mass scale (GeV) is assured. To couple the DM to the SM  we adopt the model put forward in \cite{Finkbeiner:2007kk}. Although this was originally intended as a mechanism for explaining the observed INTEGRAL/SPI positron excess \cite{Knodlseder:2005yq}, with a different choice of parameters the model can serve our astrophysical aims quite well. We begin with the introduction of a Dirac fermion comprised of the two Weyl spinors $\x_{1\cl}$ and $\x_{2\ccr}$, which couple to a real singlet scalar $\f$. The mass eigenstates for the $\x_1$ and $\x_2$ fields (which we will call $\x$ and $\x^*$, respectively) are the DM in this model. 

We impose a discrete $\mathbb{Z}_4$ symmetry under which the fields transform as
\bea
\x_{1,2}&\rightarrow& i\x_{1,2}\,,\nn\\
\f&\rightarrow&-\f\,,
\eea
but remain singlets under the symmetries of the SM. This allows for the following Lagrangian:
\bea
\mathcal{L}&=&\frac{1}{2}\partial_\m\f\partial^\m\f\,+\sum^{2}_{i}\,\x^\dagger_i\s_\m\partial^\m\x_i\,-\,m_D\overline{\x_1}_\cl\x_{2\ccr}\label{lag} \\
&-&\l_1\f\overline{\x_1}_\cl(\x_1)^c_\ccr\,-\,\l_2\f\overline{\x_2}_\ccr(\x_2)^c_\cl\,-\,V(\f,H,H^\dagger)\,+\,h.c\,.\nn
\eea
At this stage the two mass eigenstates both have mass $m_D$. To lift this degeneracy we need to break the $\mathbb{Z}_4$ symmetry. We break the symmetry spontaneously down to $\mathbb{Z}_2$ by allowing $\f$ to obtain a vacuum expectation value. The Higgs potential is given by
\bea
V(\f,H,H^\dagger)&=&\frac{\l_\f}{4}\f^4\,-\,\frac{\m_\f^2}{2}\f^2\,+\,\frac{\l_h}{4}\left(H^\dagger H\right)^2\nn\\
&-&\frac{\m_h^2}{2}H^\dagger H\,+\,\frac{\a}{2}\,\f^2\,(H^\dagger H)\,,\label{hpot}
\eea
where $H$ is the SM Higgs doublet. The last term in Eq.(\ref{hpot}) is included as it is allowed by all the symmetries of the theory. Minimizing the above potential with respect to both $\f$ and $H$ we obtain the conditions
\bea
2\l_\f\ang{\f}^2\,-\,2\m_\f^2\,+\,\a\ang{h}^2\,=\,0\,,\nn\\
\l_h\ang{h}^2\,-\,2\m_h^2\,+\,2\a\ang{\f}^2\,=\,0\,.
\label{min}
\eea
It can now clearly be seen that $\ang{\f}\neq0$. As in \cite{Finkbeiner:2007kk}, the spontaneous breaking of the discrete $\mathbb{Z}_4$ symmetry will lead to the formation of domain walls, which may be disfavored by observation. We can remove this potentially troubling phenomenon by introducing the explicit breaking term $\m \f^3$ to our Higgs potential, where $\m$ is small for reasons of technical naturalness.

Perturbing Eq.(\ref{lag}) about the vacuum there arise Majorana masses $\l_1\ang{\f}$ and $\l_2\ang{\f}$ for $\x_1$ and $\x_2$, respectively. The Lagrangian therefore contains the mass matrix
\bea
\mathcal{L}\,\supset\,-\,\left(\begin{array}{cc} \overline{\chi_{1}}_{\cl} & \left(\overline{\chi_{2}}_{\ccr}\right)^c\end{array}\right)
\left(\begin{array}{cc} \lambda_{1}\langle\phi\rangle & \frac{m_{\cd}}{2} \\ \frac{m_{\cd}}{2} & \lambda_2^*\langle\phi\rangle\end{array}\right)
\left(\begin{array}{c} \left(\chi_{1\cl}\right)^c\\\chi_{2\ccr}\end{array}\right)
\label{massmat}
\eea
which we then diagonalize to obtain the Majorana mass eigenstates $\x$ and $\x^*$:
\bea
\x&\simeq&\frac{1}{\sqrt{2}}\left[(\x_1+\x_2)^c-(\x_1+\x_2)\right]\,,\nn\\
\x^*&\simeq&\frac{1}{\sqrt{2}}\left[(\x_1+\x_2)^c+(\x_1+\x_2)\right]\,,
\label{majfields}
\eea
whose masses we find to be
\bea
m_{\x^*,\x}\,=\,\frac{1}{2}\sqrt{m_D^2+4\lnn^2\ang{\f}^2}\,\pm\,\lp\ang{\f}\,,
\eea
where $\l_\pm\equiv\frac{1}{2}(\l_1\pm\l_2^*)$. We want the mass splittings to be small, so we choose $m_D\gg\l_\pm\ang{\f}$, making $m_{\x^*,\x}\simeq\frac{m_D}{2}\pm\lp\ang{\f}$, and thus $m_{\x^*}\e=2\lp\ang{\f}$. Typically in this model we shall consider masses in the range $m_{\x,\x^*}\sim(50$-$800)$ GeV, a breaking scale of $\ang{\f}\sim(3$-$20)$ MeV, and coupling strength of $\l_\pm\sim10^{-1}$ [implying $\D m\sim(0.4$-$8)$ MeV for $\e\simeq(0.7$-$1)\times10^{-5}$]. For a detailed discussion of the parameter space, see Sec.~\ref{sec:constraints}. 

In the basis of the mass eigenstates, the Lagrangian contains the following interaction terms, which mediate both decay and scattering or annihilation processes:
\bea
\mathcal{L}&\supset&\l_+\f\overline{\x}\x\,-\,\l_+\f\overline{\x^*}\x^*\,-\,\l_-\f\overline{\x}\g_5\x^*\,\nn\\
& &-\,m_\x\overline{\x}\x\,-m_{\x^*}\overline{\x^*}\x^*\,+\,h.c\,.\label{intlag}
\eea
It should be noted that interaction terms coupling like mass eigenstates are scalar, while off-diagonal coupling is pseudoscalar. As will be seen this has a substantial effect on the DM decay rate. 

\begin{figure}[t]
\begin{center}
\includegraphics[width=5cm]{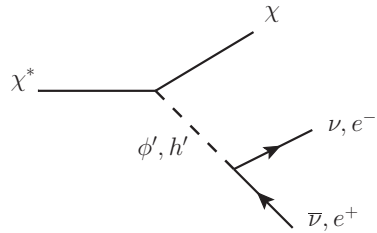}
\end{center}
\caption{Primary DM decay channel for the $\mathbb{Z}_4$ model\label{decZ4}
}
\end{figure}

Mixing of the SM sector with the dark sector ($\x$, $\x^*$, $\f$) occurs through the last term in Eq.(\ref{hpot}). Expanding the Higgs potential about the vacua of both $\f$ and $H$ produces off-diagonal mass terms for both fields. Expressing the potential in terms of the mass eigenstates $\f'$ and $h'$, we find the following mixing of states:
\bea
\phi&\simeq&\cos{\th}\,\phi'-\sin{\th}\,h'\simeq\,\phi'-\th\,h'\,,\\\nn
h&\simeq&\cos{\th}\,h'+\sin{\th}\,\phi'\simeq\,h'+\th\,\f'\,,
\label{hcomb}
\eea
where $h$ is the SM Higgs boson, and $\l_h$ is its self-coupling. To first order in $\a$
\bea
\th\,\simeq\,\frac{\alpha\langle\phi\rangle}{\l_h\langle h\rangle}\,.
\eea
and we find masses of $\f'$ and $h'$ to be
\bea
m_\f^2&\simeq&2\l_\f\ang{\f}^2\,,\nn\\ 
m_h^2&\simeq&\frac{1}{2}\l_h\ang{h}^2\,,
\eea
(in the limit $m_\f\ll m_h$) with that of $\f'$  being $\sim(2-20)$ MeV. In this work we adopt a SM Higgs mass of $m_h\sim130$ GeV. Through $\f$-Higgs mixing $\x$ and $\x^*$ can couple to $h'$ and by extension the SM. In particular, it allows the possibility of decay into SM final states via processes such as that shown in Fig.\ref{decZ4}.

This process has a decay rate given by
\bea
\G\,\simeq\,\frac{\l_-^2\,y_l^2\,\th^2}{2800\,\p^3\,m_\f^4}\,m_{\x^*}^5\e^7\,,
\label{dec1}
\eea
where $y_l$ is the Yukawa coupling for the dominant SM final state, and we assume $\D m\gg m_l$, $\lp\simeq\lnn$, and that the light final states are Dirac. Just as we choose our splitting to be small, we can choose the region of parameter space in which the lifetime is sufficiently large to disturb structure formation. 

The above decay rate contains several elements which naturally lead to suppression and thus a long lifetime. First, the decay rate is subject to phase-space suppression, as there are 3 bodies in the final state. Second, it depends on the Yukawa coupling $y_l$, which, given we are only interested in decay into light leptons ($e^+e^-$, $\overline{\n}\n$), will be a small number. It is also dependent on the Higgs mixing angle $\th$, which in turn varies depending on the strength of the coupling $\a$. Although there is some freedom of choice with respect to the value of $\a$, we typically take $\a\sim10^{-5}$, which results in a mixing of $\th\sim10^{-9}$ (see Sec.~\ref{production} for details). Lastly, pseudo-scalar coupling between $\x$ and $\x^*$ means the decay rate contains a factor $\e^7$ (as opposed to $\e^5$ for scalar coupling). The conjunction of these factors means that the DM lifetime can be long without $\l_\pm$ being too small, typically $\l_\pm\sim10^{-1}$. 

\subsection{Dark Decays $[U(1)]$}
\label{U1}

An advantage of SM final states is their detectability. While directly observable consequences are a desirable model building goal, the nonobservation of the signatures of a model can lead to constraints, as will be seen in the next section. Were observational constraints to strengthen, there is the possibility that a nonobservation of the final states in the above model may rule it out. Should this occur the viability of decaying DM in general would rely on the primary decay channel being independent of the SM. In this section we will present a model which can realize this.

One way to naturally produce a decay channel with a light final state is to upgrade the discrete $\mathbb{Z}_4$ symmetry to a global $U(1)$ symmetry. Spontaneously breaking this will produce a Nambu-Goldstone boson (NGB) in the theory, which will couple to the DM. The Lagrangian in this scenario is similar to Eq.(\ref{lag}) except that now $\f=(1/\sqrt{2})(\f_1+i\f_2)$, and $\x_{1,2}$ and $\f$ now transform under the $U(1)$ symmetry as
\begin{eqnarray}
\chi_{1,2}&\rightarrow&\chi_{1,2}\,e^{i\theta_\chi}\,,\nn\\
\phi&\rightarrow&\phi \,e^{-2i\theta_\chi}\,,
\end{eqnarray}
where $\th_\x$ is some arbitrary phase. The Higgs potential will have a similar form to Eq.(\ref{hpot}) only now $\f\neq\f^\dagger$, and $\f^2$ terms become $\f^\dagger\f$. As with the discrete case, we end up with the mass matrix in Eq.(\ref{massmat}) and subsequent mass eigenstates $\x$ and $\x^*$; only now they couple to both the mass eigenstates $\f_1'$ defined by
\bea
\phi_1&\simeq&\cos{\th_1}\,\phi_1'-\sin{\th_1}\,h'\simeq\,\phi_1'-\frac{\a\ang{h}}{\l_\f\ang{\f}}\,h'\,,\\\nn
h&\simeq&\cos{\th_1}\,h'+\sin{\th_1}\,\phi_1'\simeq\,h'+\frac{\a\ang{h}}{\l_\f\ang{\f}}\,\f_1'\,,
\label{hcomb1}
\eea
($m_\f\gg m_h$ as will be seen) and the NGB $\f_2$, i.e.
\bea
\mathcal{L}&\supset&-\frac{\lnn}{\sqrt{2}}\phi_1'\overline{\chi}\gamma_5\chi^*-\frac{i\lp}{\sqrt{2}}\phi_2\overline{\chi}\chi^*\,+\,h.c\,.
\label{int}
\eea

It should be noted that the above coupling to the NGB is scalar, which results from the fact that $\f_2$ is the imaginary component of $\f$. This means that decays into the NGB contain less $\e$ suppression (one power of $\e$) than they would were the coupling pseudoscalar ($\e^3$). 

Coupling to the NGB implies the existence of long range DM-DM interactions, which can potentially affect structure formation. To avoid the issues involved with this, we will introduce  a small soft breaking term, $\frac{\m^2}{2}(\f^2+\f^{\dagger 2})$ to the Higgs potential to explicitly break the continuous $U(1)$ symmetry down to the discrete $\mathbb{Z}_4$. This gives the NGB an $\mathcal{O}(\m)$ mass, which is naturally small.

The off-diagonal interaction term in Eq.(\ref{int}) leads to the decay channel $\x^*\rightarrow\x+\f_2$, which has the decay rate
\bea
\G\,\simeq\,\frac{\lp^2}{4\pi}m_{\chi^*}\e\,.
\label{dec2}
\eea

For the case where the primary decay of the DM was into SM final states, the reason for a long lifetime was the weak mixing with the SM and the suppression from the high power of $\e$. Should, however, the decay referred to in Eq.(\ref{dec2}) be the primary channel, there is no such suppression, and we are forced to impose the additional approximate symmetry $\l_1\simeq-\l_2^*$ to make $\lp\sim\mathcal{O}(10^{-18})$, hence achieving $\t\sim\mathcal{O}$(Gyr). As $\D m=2\lp\ang{\f}$, small $\lp$ implies a high breaking scale for reasonable values of the DM mass, roughly $\ang{\f}\sim10^{14}$ GeV. As will be discussed in \ref{U1production}, $\l_\f\sim1$ making $m_\f\sim\ang{\f}$ in this model.

This high scale has the potential to cause problems. Recall the minimization conditions in Eq.(\ref{min}). In order to reproduce the correct breaking scale for the SM Higgs, either $\a$ needs to be small enough such that $\a\ang{\f}^2$ is negligible with respect to $\m_h^2$, or we have a finely tuned scenario resembling the hierarchy problem of the SM, in which $\a\ang{\f}^2$ and $\mu_h^2$ are of similar order. The former, though the more natural of the two cases, precludes production via mixing with the SM, so we will entertain the latter for the time being.

\subsection{Production}
\label{production}
Both scenarios presented have all the required elements to disrupt structure formation in the desired fashion. All that is needed now is a production mechanism for the dark matter candidate in each scenario. As mentioned previously, one of the appealing properties of a WIMP is attainment of the correct relic abundance through thermal freeze-out from the bath in the early Universe. 

\subsubsection{$\mathbb{Z}_4$ Case}
In the model presented in \ref{Z4}, $\x$ couples to the SM through the Yukawa sector. It therefore follows that it is through these channels that it will  maintain equilibrium with the SM prior to freeze-out. Production differs from the standard WIMP scenario in that it is a two-phase process. The $\f$ is populated via interactions with the SM in the Higgs sector, while $\x$ and $\x^*$ are produced through their coupling to $\f$. At some temperature below the DM mass the $\f$-$\x$ annihilation rate will drop below the expansion rate, and $\x$ and $\x^*$ will freeze out with respect to $\f$, fixing the comoving DM abundance to the standard value. We will now chronologically step through the processes leading to DM freeze-out.

\begin{figure}[t]
\begin{center}
\includegraphics[width=5cm]{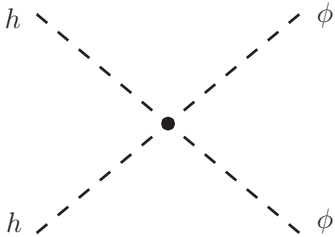}
\end{center}
\caption{Dominant $\f$ production mechanism for $T>m_h$.\label{phscat}}
\end{figure}

The requirement that $\f$ be in chemical equilibrium with the SM well before $\x$-$\f$ freeze-out places a constraint on the allowed values of the coupling $\a$. Well above the electroweak scale, the dominant process keeping $\f$ in chemical equilibrium with the SM will be $hh\rightarrow\f\f$ (Fig.\ref{phscat}), which at temperatures well above the Higgs mass has the annihilation rate
\bea
\G(hh\rightarrow\f\f)\simeq\frac{\a^2 \,T}{256\p^3}\,.
\label{hotphi}
\eea

This process will remain in equilibrium until $T <m_h$, and $h$ production becomes Boltzmann suppressed, causing this $\f$ production channel to become unavailable. For the DM masses of interest in this model (of order or below $m_h$), we require $\a > 10^{-6}$ to ensure that $\f$ is in equilibrium at some point prior to $\f$-$\x$ freeze-out.

Also contributing to $\f$ production is the $h$-mediated processes $\overline{f}f\rightarrow\f\f$, which have the annihilation rate of
\bea
\G(f\overline{f}\rightarrow\f\f)\simeq\frac{\a^2\, y_f^2\,\ang{h}^2\,T^3}{16\p^3\,m_h^4}\,.
\label{hotphi1}
\eea
 up to a color factor for processes involving quarks. The temperatures at which these processes freeze out depend on both the Higgs mixing $\a$ and SM fermion Yukawa $y_f$. For the values of $\a$ considered in this paper ($\a\sim10^{-5}$) the process which remains in equilibrium longest is that involving $b$ quarks. This freezes out around the time at which the annihilation in Fig.\ref{phscat} turns off. Thus the temperature at which $\f$ freezes out with respect to the SM can be calculated to be $T_f^{\f\mhyphen SM}\sim20$ GeV. This occurs when $\f$ is still relativistic.

\begin{figure}[t]
\begin{center}
\includegraphics[width=5cm]{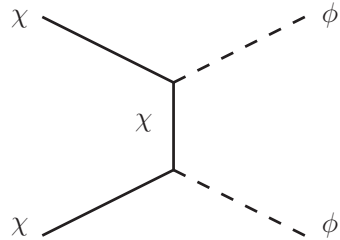}
\end{center}
\caption{DM produced through $\f$ annihilation.\label{DMpscat}}
\end{figure}

After $\f$-$SM$ freeze out, the temperature of the $\f$-$\x$ system will continue to track that of the background.\footnote{Up to a factor $(g_*'/g_*)^{1/3}$, where $g_*$ and $g_*'$ are measures of the number of freedom in the bath at $T_f^{\f\mhyphen SM}$ and $T_f^{\f\mhyphen\x}$  respectively. We will assume this ratio to be $\sim1$, and $\sqrt{g_*}\simeq10.8$, i.e. that all degrees of freedom are in equilibrium. While depending on the time of DM freeze out this might not be strictly true, the effect on the results will be negligible. It is therefore irrelevant exactly when $\f$ freezes out with respect to the SM, as long as it has been in equilibrium at some point prior to $\f$-$\x$ freeze out.} The DM will be kept in chemical equilibrium with $\f$ through the scattering in Fig.\ref{DMpscat}, which in the nonrelativistic limit has a cross section of
\bea
\s\,v_{rel}\simeq \frac{|\lp|^4}{\p\,m_\x^2}\label{cross}\,.
\eea
This process will freeze out once the temperature of the $\f$-$\x$ system falls below $m_\x$ and the number density of $\x$ becomes Boltzmann suppressed.

To determine the DM relic abundance we use the well established result~\cite{Kolb:1990vq}
\bea
\W_\x h^2&=&1.07\times10^9\frac{x_{\mathit{DM\mhyphen f}} \sqrt{g_*}GeV^{-1}}{g_{*s}m_{Pl}\ang{\s\,v}}\,,
\label{relic}
\eea
where 
\bea
x_{DM\mhyphen f}&=&m_\x/T_f^{\f\mhyphen\x}\,=\,\ln[0.038(g/\sqrt{g_*})m_{Pl}m_\x\ang{\s\,v}]\nn\\
&&-\frac{1}{2}\ln\left[\ln[0.038(g/\sqrt{g_*})m_{Pl}m_\x\ang{\s\,v}]\right]\,,
\label{xf}
\eea
and $T_f^{\f\mhyphen\x}$ is the temperature at which $\f$ and $\x$ drop out of chemical equilibrium. We find that typically $x_{DM\mhyphen f}\sim20$. The requirement that we produce the observed relic density of $\W_\x\simeq0.22$, places constraint on the free parameters $\lp$ and $m_\x$ should the DM be a thermal relic. See Sec.~\ref{sec:constraints} for a full treatment of the parameter space. Given that $m_\x\simeq m_{\x^*}$ (and $\D m\ll T_f^{\f\mhyphen\x}$) $\x$ and $\x^*$ will be produced in equal abundance.

\begin{figure}[t]
\begin{center}
\includegraphics[width=5cm]{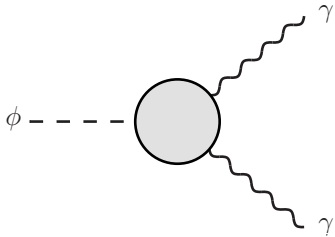}
\end{center}
\caption{1-loop order decay $\f\rightarrow\g\g$, through $h$-$\f$ mixing. Includes contribution from loops involving $W^\pm$, unphysical charged Higgs components $h^\pm$, and Fadeev-Popov ghosts.\label{phidecdiag}}
\end{figure}

After $\f$-$\x$ freeze-out, the relativistic $\f$ will remain with fixed abundance until the spontaneous breaking of the $\mathbb{Z}_4$
symmetry (at MeV scale). After symmetry breaking they become unstable to decay into photons via the loop order process depicted in Fig.~\ref{phidecdiag} \cite{Djouadi:2005gj,Ellis:1975ap}. This process has a rate of
\bea
\G(\f\rightarrow\g\g)&\simeq&\frac{G_F\,\a_{EM}^2\,\th^2\,M_W^4}{2\sqrt{2}\p^3\,m_\f}\\
&\simeq&4.5\times10^{4}s^{-1}\left(\frac{\th}{10^{-9}}\right)^2\left(\frac{10 MeV}{m_\f}\right),\nn
\label{phidec}
\eea
which is large compared to the expansion rate, and $\f$ rapidly depopulates.

In our calculation of the process depicted at tree level in Fig.~\ref{DMpscat}, we have omitted the contribution from ladder diagrams involving $\f$ exchange in the initial state. This approximation is valid at high energies, but begins to break down near freeze out, when the DM is in the moderate-nonrelativistic regime. 

At low velocity the Yukawa potential (resulting from $\f$ exchange) from one initial state $\x$ can significantly distort the wave-function of the other from that of a free particle. This leads to an enhancement of the velocity averaged cross section in an effect known as Sommerfeld enhancement \cite{ArkaniHamed:2008qn,Feng:2010zp,Cassel:2009wt,Slatyer:2009vg}. This effect can be taken into account by multiplying the relevant cross section by a velocity dependent Sommerfeld factor $S$. To calculate the enhancement to the process in Fig.~\ref{DMpscat} we follow the method of \cite{Feng:2010zp,Cassel:2009wt}, but find that in the relevant region of parameter space $S$ is close to 1 and the enhancement negligible. The enhancement generally becomes more important for larger values of $\D m$.

\subsubsection{$U(1)$ Case}
\label{U1production}

Production in the second model presented is slightly more difficult. An unfortunate consequence of a small Yukawa coupling is a weakening of the annihilation cross section (Fig.~\ref{DMpscat}). This suppression ensures that the process in Fig.~\ref{DMpscat} is never in equilibrium, making thermal production of the DM impossible. This leads us to consider a nonthermal production mechanism, in which $\x$ and $\x^*$ are produced out of equilibrium through their weak mixing with the bath. Another possibility is production through direct coupling of the DM to the inflaton. While this is clean in that it is independent of SM processes, it requires fine tuning to attain the correct relic abundance. For the time being we will entertain the former possibility. 

\begin{figure}[t]
\begin{center}
\includegraphics[width=5cm]{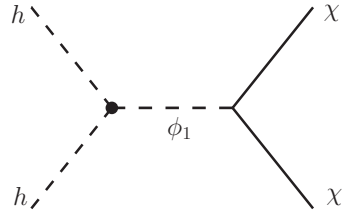}
\end{center}
\caption{Dominant DM production mechanism in the $U(1)$ model. \label{hDMscat}}
\end{figure}

The dominant channel through which production can occur is through the SM Higgs annihilation pictured in Fig.~\ref{hDMscat}. As $\ang{\f}$ is large in this model, we expect this process to be strongest above the electroweak breaking scale. At these high temperatures finite temperature effects come into the Higgs potential at loop order \cite{Parwani:1991gq,Dolan:1973qd,Weldon:1983jn}. This has the effect of giving the scalar components of the SM Higgs doublet temperature-dependent masses of the form
\bea
m_h^2\simeq\frac{\l_h T^2}{24}\,.
\eea
The process in Fig.~\ref{hDMscat} goes to a maximum near the $\f_1'$ resonance, in which region $m_{\f_1}^2\simeq4m_h^2$. Granted $\a^2\ll\l_\f$, and following the analysis of \cite{Gondolo:1990dk}, the velocity averaged cross section can in this region be well approximated by
\bea
&&\langle\sigma v\rangle\,\simeq\, \frac{\lambda_+^2Tm_\phi^3}{\pi^3(n_h^0)^2}K_1\left(\frac{m_\phi}{T}\right)\times\label{crossnt}\\
& &\frac{\frac{\alpha^2}{\l_\f}\sqrt{m_\phi^2-4m_h^2}}{\left[\left(\frac{\alpha^2}{\l_\f}\sqrt{m_\phi^2-4m_h^2}+\l_\f m_\phi\right)\coth{\left(\frac{m_\f}{4T}\right)}+128\lambda_+^2m_\phi\right]}\,.\nn
\eea

To avoid Boltzmann suppression in Eq.(~\ref{crossnt}) we will take $\l_\f$ to be small for now ($\l_\f\sim10^{-17}$), which implies that $m_{\f_1}$ is far below the $U(1)$ breaking scale ($m_{\f_1}\sim 100$ TeV). 
In order to calculate the  abundance at a particular temperature, we must solve the comoving Boltzmann equation, which can be expressed in the form
\bea
\frac{dn_\chi(T)}{dT}-\frac{3}{T}n_\chi(T)=-\frac{(n_h^0)^2}{HT}\langle\sigma v\rangle\,,
\eea
and has the solution
\bea
n_\x(T)\,=\,T^3\int^{T_{n_\x=0}}_{T}\frac{(n_h^0)^2\langle\sigma v\rangle}{HT'}dT'\,,\label{noint}
\eea
where $T_{n_\x=0}\simeq\sqrt{6/\l_h}$ is defined by the temperature at which $m_{\f_1}^2\simeq4m_h^2$ and will be taken to be when significant production starts.

For the representative region of parameter space $\l_\f\sim10^{-17}$, $\a\sim10^{-15}$, $\lp\sim10^{-18}$, and $\t_{\x^*}\sim1$ Gyr, the comoving number density can be calculated to be $\mathcal{O}(10^{-27})$, roughly 17 orders of magnitude below the required value at that temperature. These values for the parameters in the model were chosen as they were shown to maximize production. As this channel is expected to be the strongest available it is therefore clear that production of the DM via mixing with the SM in such a model is impossible. The implication of neither a SM final state nor SM related production is the independence of the dark sector from the visible. This gives us complete freedom in the choice of dimensionless parameters $\a$ and $\l_\f$ but precludes entirely the possibility of direct verification of the model. We can now choose $\l_\f\sim1$ and $\a$ to be very small to avoid issues of fine-tuning.

Independence of the dark sector from the SM implies the necessity for some novel DM production mechanism. As mentioned earlier this can be realized through a direct coupling of the DM to the inflaton, but as stated such a mechanism is problematic as it is difficult to obtain a relic density of order that of the SM without fine-tuning of the DM-inflaton coupling.

\subsection{Depopulation of the Excited State}
\label{depopulation}

\begin{figure}[t]
\begin{center}
\includegraphics[width=5cm]{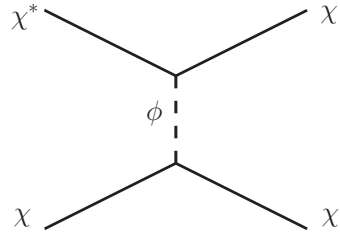}
\end{center}
\caption{Process by which $\x$ and $\x^*$ maintain chemical equilibrium.\label{dwnscat}}
\end{figure}

In the $\mathbb{Z}_4$ model, as the temperature of the $\f$-$\x$ system drops well below the DM mass, $\x$ and $\x^*$ will have chemically frozen out fixing the relic abundance. The s-channel equivalent to Fig.~\ref{DMpscat} will, however, maintain kinetic equilibrium in the $\f$-$\x$ system to temperatures down as low as $m_\f$ \cite{Finkbeiner:2008gw,Feng:2010zp}. Both $\x$ and $\x^*$ will be kept in equilibrium with each other by way of a process like that in Fig.~\ref{dwnscat}, causing both to track closely the temperature of the background. However, as the average kinetic energy drops below $\D m$ the process $\x\x\rightarrow\x^*\x$ is no longer kinematically viable, and the up-scattering rate becomes Boltzmann suppressed \cite{Finkbeiner:2008gw}. The result is a rapid depopulation of $\x^*$, and an absence of the heavy state so important for disturbance of structure formation. This issue can be averted should the scattering rate for Fig.~\ref{dwnscat} be small enough such that the process freezes out sufficiently early, i.e for $T\gg\D m$. Should this be the case, both the forward and back scattering processes will cease well before depopulation becomes an issue.

The cross section for this process (at tree level) can be calculated to be
\bea
\s v_{rel}&\simeq&\frac{3|\l_-\l_+|^2}{\p m_{\x^*}^2}\frac{1}{v_{rel}}\log{\left[\frac{32}{v_{rel}^2}\right]}\,,\label{cross2}
\eea
in the limit $m_\f^2\ll m_{\x^*}\D m$, which is justified in the region of parameter space considered (see Sec.~\ref{sec:constraints}). In the moderate to nonrelativistic regime, the scattering rate for process $\x^*\x\rightarrow\x\x$ is given by
\bea
\G\,\simeq\,(n_\x)\,\frac{x_{sc\mhyphen f}^{3/2}}{2\sqrt{\p}}\int^{1}_{0}(\s v_{rel})\,S\, v_{rel}^2\,e^{-x_{sc\mhyphen f} v_{rel}^2 /4}\,dv_{rel},
\eea
where $x_{sc\mhyphen f}=m_\x/T_{f}^{\x\mhyphen\x^*}$, and $T_{f}^{\x\mhyphen\x^*}$ is defined as the temperature at which the process in Fig.~\ref{dwnscat} freezes out.

After the process in Fig.~\ref{DMpscat} freezes out, the comoving DM
number density $n_\x/T^3$ is fixed, and is given by \bea
n_\x/T^3&\simeq&\frac{g_{*s}}{m_\chi}\,3.76\times10^{-11}\,GeV\,.\nn
\eea where $T$ is the temperature of the bath. We can now choose
parameters such that the process in Fig.~\ref{dwnscat} freezes out
around the same time as that of Fig.~\ref{DMpscat}, in which case
Eq.(\ref{cross}) and Eq.(\ref{cross2}) are of a similar order. In the
relevant region of parameter space, this is generally the case, with
$x_{sc\mhyphen f}\sim1$. Interestingly, this is before $\f$-$\x$
freeze-out, meaning $\x$ and $\x^*$ are both are in equilibrium with
$\f$ but not each other. 

In the above, we have considered only the depopulation of $\x^*$ in the
early Universe.  It is also important that $\x^*$ not be depopulated
via scattering in the late Universe, when Sommerfeld effects are
significant.  In fact, additional constraints on $\x\x$ and $\x\x^*$
scattering arise from the requirement that self-scattering of DM does
not significantly perturb galactic halo shapes~\cite{Feng:2009hw}.
These requirements will be taken into account in
Sec.~\ref{sec:constraints}.

In the $U(1)$ model there are no such depopulation issues, as $\lp$ is very small and the DM is never in equilibrium in the first place.

\section{Constraints on $\mathbb{Z}_4$ model}
\label{sec:constraints}

\begin{figure}[t!]
\begin{center}
\includegraphics[width=8cm]{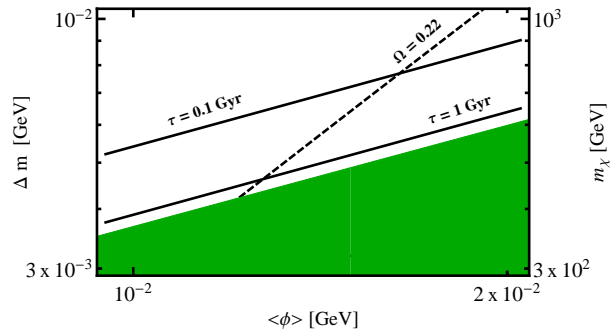}
\end{center}
\caption{Available parameter space for decays into $e^+ e^-$ ($y_l=y_e$) for $\l_\f=1$ and $\a=10^{-5}$ for lifetimes $\t_{\x^*}=0.1$ Gyr  (solid black upper line), and $\t_{\x^*}=1$ Gyr (solid black lower line). Parameters yielding correct freeze-out abundance lie on the dashed black line. Shaded is the exclusion region from \cite{Bell:2010fk}. We have chosen $\e=10^{-5}$.}
\label{epemparam}
\end{figure}

\begin{figure}[t!]
\begin{center}
\includegraphics[width=8cm]{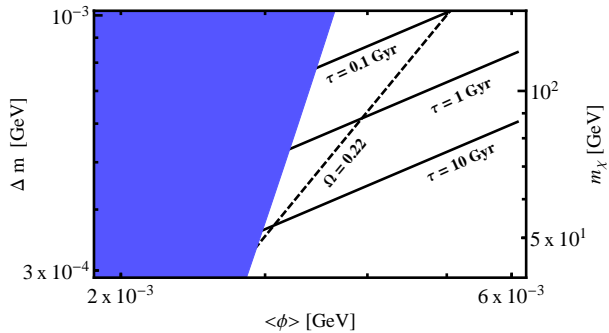}
\end{center}
\caption{Available parameter space for decays into $\ov{\n}\n$ with a larger Yukawa of $y_\n\simeq10^{-4}$ ($\n$ are Majorana) for $\l_\f=0.8$ and $\a=10^{-5}$ for lifetimes $\t_{\x^*}=0.1$ Gyr (solid black upper line), $\t_{\x^*}=1$ Gyr (solid black center line) and $\t_{\x^*}=10$ Gyr (solid black lower line). Parameters yielding correct freeze-out abundance lie on the dashed black line.  Shaded is the exclusion region based on ellipticity constraints \cite{Feng:2009hw}. We have chosen $\e=0.7\times10^{-5}$.}
\label{nuMajparam}
\end{figure}

Up to this point there has been minimal discussion of the choice of values for the many free parameters in our model. In order to do so clearly it is important to understand exactly what constraints are present. There are initially 7 independent free parameters, those related to the fermions $\x$ and $\x^*$, namely, $m_{\x^*}$ and $\l_\pm$, and those belonging to the Higgs sector: $\l_\f$, $\a$, $\m_\f$, and $\m$. Recall also that we can express the mass splitting in terms of these parameters, that is, $\D m=2\lp\ang{\f}$ [$\ang{\f}$ depends on Higgs potential parameters from Eq.(\ref{min})]. Thus when we parametrize $\D m$ in terms of $\e$ ($\D m=m_{\x^*}\e$) and fix its value to $\e=10^{-5}$, we place a constraining relationship between $\l_+$, $m_{\x^*}$, and $\ang{\f}$. As a second constraint we will impose $\lp\sim\lnn$, as they will only differ greatly in the finely tuned scenario where $\l_1\simeq\pm\l_2$ to high precision.  Last, we must satisfy the condition in Eq.(\ref{relic}), to ensure correct relic abundance. These three constraints reduce the number  of free independent parameters to 4, which can be taken to be $\ang{\f}$, $\l_\f$, $\a$, and $\m$. We can now express allowed values of $\D m$ as a function of breaking scale $\ang{\f}$ for chosen vales of $\a$ and $\l_\f$. We must choose $\a$ appropriately such that $\f$ goes into equilibrium with the bath before the temperature of DM freeze-out. The allowed values of $\D m$ for the appropriate DM lifetimes are plotted on the left hand side of Figs.~\ref{epemparam}-~\ref{nuMajparam}, while the corresponding values of $m_{\x^*}$ (for $\e=10^{-5}$ and $\e=0.7\times10^{-5}$ for Figs.~\ref{epemparam} and ~\ref{nuMajparam}, respectively) are plotted on the right.

The presence of readily detectable charged particles in the final state increases the possibility of both direct and indirect detection. Indeed heavy constraints can be placed on the parameter space based on nonobservation of the consequences of such a final state. In \cite{Yuksel:2007dr,Bell:2010fk} detailed analyses of the photon, positron and neutrino backgrounds were performed with decaying DM models in mind, and constraints placed on the relevant parameters $\t_{DM}$ and $\D m$. We have translated the constraints on decay to $e^+e^-$ to the parameter space relevant to this model, resulting in the exclusion region in Fig.~\ref{epemparam}.

Which leptons will be produced predominantly will depend not only on the choice of parameters (i.e. for which lepton does $\D m \geq2m_l$ hold) but also the choice of neutrino model. We will consider three distinct possible final states: (i) $e^+e^-$, i.e. $\D m\geq2m_e$ for Dirac neutrinos, (ii) $\n\overline{\n}$, i.e. $\D m<2m_e$ for Dirac neutrinos, and (iii) $\n\overline{\n}$ for Majorana neutrinos.

(i) If we consider the SM neutrino to be a Dirac particle, then the upper bound on light neutrino masses implies a Yukawa coupling of $y_\n\lesssim10^{-11}$ \cite{Fukugita:2003en}. Thus when decays into charged leptons are kinematically allowed ($\D m\geq2m_e$), their relatively large coupling in the Yukawa sector will render decays into neutrinos subdominant. There is the important constraint that $\D m < 2m_\m$, as should $\m^+\m^-$ pairs be produced, their Yukawa is large enough that for no allowed values of $\D m$ and $\ang{\f}$ would $\t_{\x^*} > $0.1 Gyr. Thus for $\D m\geq2m_e$, decays to $e^+e^-$ will dominate. A representative region of parameter space can be seen in Fig.~\ref{epemparam}. To obtain the correct relic abundance, parameters must lie on the dashed line. We find that for $m_\x\sim$600 GeV, the breaking scale $\ang{\f}$ is required to be in the $\sim$10 MeV range, while (for $\e=10^{-5}$) $\D m$ is in the MeV. It should be noted that these parameters coincide with an $x_{sc\mhyphen f}\sim1$, which is well above $T_{f}^{\x\mhyphen\x^*}\sim\D m$, removing the possibility of depopulation of the heavier DM state.

Interestingly, the (1-30) Gyr lifetime range preferred by Abdelqader and Melia \cite{Melia} has been nearly completely excluded for decays into charged particles, leaving only the restrictive region of (0.1-1) Gyr available. It should be noted however that should decays to neutrinos dominate, we can avoid this exclusion region entirely.

(ii) Should $\D m < 2m_e$  only neutrinos are kinematically available. As Dirac neutrinos couple only very weakly with the SM Higgs, the lifetime of the DM will be too long to affect structure formation. There are two ways in which we could reduce $\t_{\x^*}$, by either increasing $\a$ or decreasing $m_\f$. We find, however, that for $m_\f >\D m$, there are no values of $\a$ and $m_\f$ that can yield a lifetime short enough. If, however, $m_\f\lesssim\D m$, the process $\x^*\rightarrow\x+\f'$ becomes kinematically allowed. The rate for this process does not contain the high level of suppression that decays into SM final states suffer, and we find its lifetime to be $\ll0.1$ Gyr, dominating over decays into $\n\overline{\n}$. Thus for the choice of parameters $m_\f >\D m$ the DM lifetime is too long, and for $m_\f\lesssim\D m$ $\n\overline{\n}$ final states are unimportant, and $\t_{\x^*}$ is far too short. It therefore seems that in no region of parameter space can decays into Dirac neutrinos affect structure formation.

(iii) Should we introduce Majorana masses for the $\n_\ccr$ and employ the type I see-saw mechanism, we have the freedom to make $y_\n$ large enough (while still keeping the neutrino mass small) such that decays with neutrino final states will dominate without the need for fine-tuning $m_\f$. We can consider three options: $y_\n^2 \ll y_e^2$, $y_\n^2\simeq y_e^2$, and $y_\n^2 \gg y_e^2$. Should $y_\n^2 \simeq y_e^2$ or $y_\n^2\ll y_e^2$, decays into electrons are either important or dominate, and so allowed parameters will be the same as for the Dirac case. However, for $y_\n^2\gg y_e^2$, neutrino final states are dominant for all values of $\D m$, and while we still need to respect the observational constraints in Fig.~\ref{epemparam}, we have a wider parameter space available, an example of which can be seen in Fig.~\ref{nuMajparam}.

Conversely to before, having $\D m\geq2m_e$ makes $\t_{\x^*} > $ 0.1 Gyr impossible, as the Yukawa controlling the decay is much larger than that of the electron. This constraint requires us to choose much smaller values of $m_\f$ than for $e^{+}e^{-}$ final states, if we wish to maintain thermal production. The smaller the value of $m_\f$, the larger the $\x$-$\x$ cross section in present-day halos. In \cite{Feng:2009hw} authors argued that to maintain the observed ellipticity in galactic halos, the timescale for DM self-interactions must be longer than the halo age ($\G_{DM-DM}^{-1}>10^{10}$ yr). Following the approach in \cite{Feng:2009hw} constraints were placed on our parameter space, resulting in the shaded exclusion region in Fig.~\ref{nuMajparam}.

\section{Conclusions}
\label{sec:conclude}

Models for decaying dark matter are interesting in that they maintain
the attractive features of the $\Lambda$CDM model, while alleviating
the issues pertaining to the over prediction of small scale power. In
this work we investigated two examples of the class of DM models in
which decay occurs via the process $\x^*\rightarrow\x+l$, where $\x^*$
and $\x$ are nearly degenerate in mass (in this work we chose $\D
m/m_{\x^*} \equiv \epsilon \simeq 10^{-5}$) and $l$ is relativistic.

In the first scenario, we considered the possibility of decays into SM
final states. We demonstrated that through the breaking of a discrete
$\mathbb{Z}_4$ symmetry with the real scalar field $\f$, we could both
produce two Majorana DM candidates $\x^*$ and $\x$ with nondegenerate
mass, and allow for the decay channel $\x^*\rightarrow\x+\textrm{SM}$.
The required long lifetime [(0.1-100) Gyr] was naturally achieved, as
the the decay rate was suppressed by a high power of $\epsilon$, by small Yukawa
couplings, and by the small mixing between SM-sector and dark-sector
particles.

The only two viable decay modes involving SM final states were
$\x^*\rightarrow\x+e^+e^-$ and $\x^*\rightarrow\x+\overline{\n}\n$,
where the latter is possible only in the case of Majorana neutrinos.
We found that for DM masses in the range (50-$800$) GeV [$\D
m\simeq(0.4-8)$ MeV] and for a $\f$-Higgs mixing of $\a\simeq10^{-5}$,
all required criteria, including thermal production, could be met if
the $\mathbb{Z}_4$ symmetry was broken at the MeV scale, with
$\ang{\f}\simeq(3-20)$ MeV ($m_\f\simeq(2-20)$ MeV).  Interestingly, in applying the
constraints on decays to $e^+e^-$ derived in \cite{Bell:2010fk}, we
showed that this final state is almost excluded for DM lifetimes in
the (1-30) Gyr range preferred in \cite{Melia}.  Dirac neutrinos were
unable to fulfill the requirements for decaying DM, as their Yukawa
coupling is too small.  Thus decays to Majorana neutrinos are
preferred by such DM decay models, as they are not constrained to
either the short lifetimes applicable for $e^+e^-$ decays nor the
small Yukawa couplings of Dirac neutrinos.

In the second scenario, we considered the possibility of non-SM final
states. This was achieved by replacing the discrete $\mathbb{Z}_4$
symmetry with a continuous $U(1)$ symmetry. Breaking of the $U(1)$
symmetry led to a pseudo-Nambu-Goldstone boson, which became the light
final state produced in decays.  As the DM decay process was no longer
strongly suppressed, we were forced to finely tune model parameters to
obtain a DM lifetime in the correct range.  A consequence of this fine-tuning was to make DM production via mixing with the SM no longer
possible.  In this scenario, the dark and visible sectors are almost
decoupled from each other.  Though aesthetically less appealing, this
model demonstrates the feasibility of decaying dark matter,
independent of the strength of coupling to the SM.

\bigskip

\section*{Acknowledgements}

NFB and RRV were supported, in part, by the Australian Research
Council, and AJG by the Commonwealth of Australia.  We thank F. Melia and M. Drewes for useful discussions, and K. Petraki for a detailed reading of the manuscript.

\bibliographystyle{h-physrev5}
\bibliography{decaymodel.bib}

\end{document}